\documentclass[fleqn,usenatbib]{mnras}
\usepackage{newtxtext,newtxmath}
\usepackage[T1]{fontenc}
\usepackage{ae,aecompl}
\usepackage{graphicx}	% Including figure files
\usepackage{amsmath}	% Advanced maths commands
\usepackage{amssymb}	% Extra maths symbols

\newcommand{\unit}[1]{\ensuremath{\, \mathrm{#1}}}  % Shorthand for proper spacing with units.

\title[Dynamical Interactions in GJ4276]{Dynamical Interactions in the Planetary System GJ4276}

\author[Horrobin \& Rein]{
Fergus Horrobin,$^{1,2}$\thanks{horrobin@astro.utoronto.ca}
Hanno Rein$^{1,2,3}$
\\
$^{1}$Department of Physical \& Environmental Sciences, University of Toronto at Scarborough, Toronto, ON M1C 1A4, Canada\\
$^{2}$Department of Astronomy \& Astrophysics, University of Toronto, Toronto, ON M5S 3H4, Canada\\
$^{3}$Centre for Planetary Sciences, University of Toronto at Scarborough, Toronto, Ontario M1C 1A4, Canada
}

\date{Accepted XXX. Received YYY; in original form ZZZ}

\pubyear{2019}

\begin{document}
\label{firstpage}
\pagerange{\pageref{firstpage}--\pageref{lastpage}}
\maketitle

\begin{abstract}
GJ4276 is an M4.0 dwarf star with an inferred Neptune mass planet from radial velocity (RV) observations.
We reanalyse the RV data for this system and focus on the possibility of a second, super earth mass, planet.
We compute the timescale for fast resonant librations in the eccentricity to be $\sim 2\,000 \unit{days}$.
Given that the observations were taken over $700\unit{days}$, we expect to see the effect of these librations in the observations.
We perform a fully dynamical fit to test this hypothesis.
Similar to previous results, we determine that the data could be fit by two planets in a 2:1 mean motion resonance.
However, we also find solutions near the 5:4 mean motion resonance which are not present when planet-planet interactions are ignored.
Using the MEGNO indicator, we analyze the stability of the system and find that our solutions lie in a stable region of parameter space.
We also find that though out of resonance solutions are possible, the system favours a configuration which is in a first order mean motion resonance.
The existence of mean motion resonances has important implications in many planet formation theories.
Although we do not attempt to distinguish between the one and two planet models in this work, in either case, the predicted orbital parameters are interesting enough to merit further study.
Future observations should be able to distinguish between the different scenarios within the next 5~years.
\end{abstract}


%%%%%%%%%%%%%%%%% BODY OF PAPER %%%%%%%%%%%%%%%%%%

\section{Introduction}

Of the recently discovered planetary systems, many are found to be in multiple planet systems. 
M dwarf stars pose a particularly interesting target since they show a relatively high planet occurrence rate of $2.5 \pm 0.2 \unit{planets}$ for 1--4 Earth radii planets with orbital periods less than $200 \unit{days}$ \citep{Dressing2015}.
It has also been predicted that M dwarfs constitute roughly 75\% of the stars in our nearby solar neighbourhood \citep{Henry2006}.

When compared to solar like stars, M dwarfs are smaller in radius, in mass and in luminosity.
This makes them favourable for the detection of exoplanets by radial velocity (RV) or transit methods since the reflex motion of a star scales with the mass of the star as $M_{\star}^{-2/3}$ \citep*{Cumming2002} and the transit depth with the radius of the star as $R_{\star}^{-2}$ \citep{Seager2003}.
Since by nature, M dwarfs have a lower luminosity, the habitable zone is moved closer to the planet than solar type stars.
Therefore, M dwarfs are a particularly favourable target for finding Earth mass exoplanets in the habitable zone.
The CARMENES survey \citep{Moreno-Raya2018} has as a main objective to search for and classify low mass planets around $\sim 300$ M dwarf stars \citep{Reiners2018}.
The survey's spectrograph has an RV accuracy of $\sim 1 \unit{ms^{-1}}$ and has been used for the discovery of several planetary systems, including GJ4276. 

\citet{Nagel2019} found that the signal in the RV data for GJ4276 is not likely related to stellar activity and is likely of planetary origin.
They fit the signal using both a one planet and two planet model.
In the one planet model, they found that the best result is achieved with an eccentricity of $0.37$.
This would be one of the most eccentric exoplanets around an M dwarf with such a short orbital period currently known.
They also considered a two planet model, however, in this case the period ratio is fixed to be in a 2:1 mean motion resonance and the eccentricities are forced to 0.

Mean motion resonances are found in many planetary systems and~2:1 mean motion resonances are particularly common.
There are two known systems with 2:1 mean motion resonances and M-stars as hosts, GJ876 \citep{Marcy2001} and TRAPPIST-1 \citep{Gillon2017}.
Mean motion resonances are dynamically interesting because their existence lets us draw conclusions about the system's formation phase. 
It is predicted that systems that are in first order resonances got there by early migration prior to disk dispersal \citep{Correia2017}.
Since disk dispersal is estimated to happen on a 1--10 Myr timescale \citep{Alexander2014}, and planet migration on a similar timescale \citep{Armitage2005}, the planets would need to form quite early in order to migrate before the disk disperses.
This in turn means that the core of the planet should have formed with plenty of gas still left in the disk to accrete an atmosphere \citep{Lissauer1993}.

We reanalyse the RV signal for the system GJ4276 in this paper.
In contrast to \citet{Nagel2019}, we perform a fully dynamical model which allows the planets to interact rather than being on Keplerian orbits.
We also allow for more freedom in our model parameters, specifically, we do not force a 2:1 period ratio in the two planet model.
Though \citet{Nagel2019} slightly favour the single planet eccentric model over the two planet solution, we remark that there is inadequate statistical difference to be able to distinguish between them and attempting to do so is beyond the scope of this work.
We focus on the two planet model since the predicted 2:1 mean motion resonance is dynamically interesting.
Given the analysis of the one planet system is not discussed in this work, we refer the reader to the original discovery paper by \citet{Nagel2019}.

We analyze the resonant behaviour and estimate the dynamical timescale in order to determine whether we could observe planet-planet interactions.
To discuss the stability of the system, we perform $N$-body integrations for $\sim 10^6 \unit{orbits}$ and map the stability of the parameter space by determining the mean exponential growth of nearby orbits (MEGNO) parameter.

\section{Fitting RV Data} \label{sec:fit_rv}

\subsection{Numerical Setup}

We use the affine invariant Markov chain Monte Carlo (MCMC) sampler \texttt{EMCEE} \citep{Foreman2013}, which is supplied as a freely available python package to sample the posterior of the orbital parameters.
We couple the MCMC sampler with the high order $N$-body integrator \texttt{IAS15} \citep{ReinSpiegel2015}, which is part of the freely available python package \texttt{REBOUND} \citep{ReinLiu2012}.
This setup allows us to simulate the reflex motion of the star and compare the results with the observed RV data.
The advantage of this method, compared to the procedure in \citet{Nagel2019} is that our model fully captures the dynamical interactions of the planets.

For each of the planets, we let the minimal mass ($m \sin i$) as well as the 4 orbital parameters $a, e, T, \omega$ (semi-major axis, eccentricity, time of pericenter passage and periastron) freely vary.
For this work we mostly fix the inclination of the system at $90^\circ$, and fix the mutual inclination at $0^\circ$.
To constrain the inclination we simulate the effect of varying the inclination of the system.
To parameterize the eccentricity, we define $h = e \sin \omega$ and $k = e \cos \omega$ such that $e = \sqrt{h^2 + k^2}$ and $\omega = \tan^{-1} \frac{h}{k}$.
This avoids the singularity at $e=0$ and speeds up convergence as described in \citet{Hou2012}.
We also allow the stellar mass $M_{\star}$ to vary freely.

To account for additional stochastic noise in the data, we follow the standard procedure of \citet{Baluev2009} and introduce a jitter parameter, $\sigma_{\rm jitter}$.
We also introduce a parameter $\gamma$ to account for the center of mass motion of the system.
These two additional parameters are fit simultaneously with the orbital parameters in the MCMC sampler.

To remain consistent with previous results, we assume uniform priors for all parameters other than stellar mass.
For stellar mass, we follow \citet{Nagel2019} and assume a Gaussian prior with mean and variance $0.406 \pm 0.030 M_{\sun}$.
For the free parameters with uniform priors, we find that the final results are insensitive to the width of the prior.
We therefore simply choose a width that captures the full dynamical range of the free parameters with a relatively fast convergence time. 

\begin{figure}
    \centering
    \includegraphics[width=\linewidth]{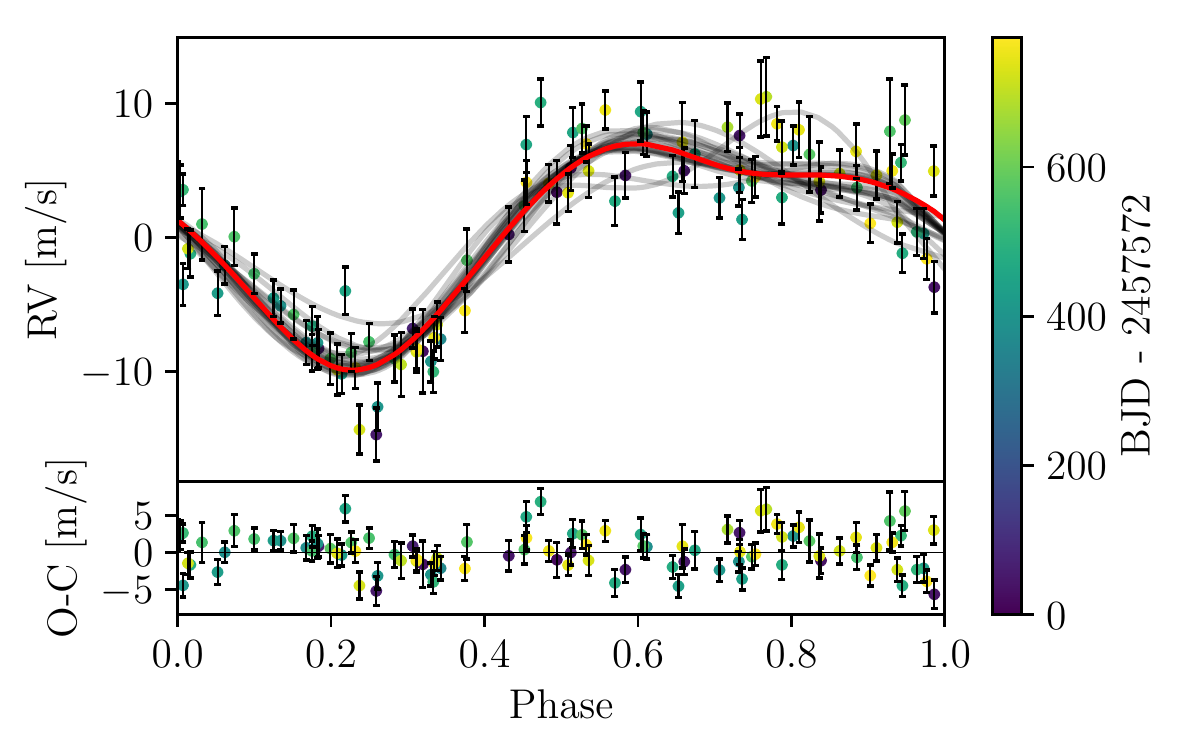}
    \caption{Radial velocity data and model. 
    The data is phase folded with period $13.347 \unit{days}$. 
    Top panel: RV data points, MCMC mean (red), MCMC samples (gray).
    Bottom panel: residual errors not including jitter, $\sigma_{\rm jitter}$.
    }
    \label{fig:rv_fit}
\end{figure}

We initialize the parameters to the values determined in \citet{Nagel2019} to speed up convergence.
Experimentation shows that the initial choice of values has little effect on the outcome if the MCMC sampler is evolved for a sufficient number of burn-in generations.
A total of 200 random walkers are initially evolved for a few $\sim 3000$ burn in steps.
The parameters are then sampled for 1000 generations of the walkers, from which we calculate the mean and $1\sigma$ confidence interval for each parameter.

\subsection{Best Fit Parameters}

Our estimated nominal best fit parameters for each planet are presented in \autoref{tab:best_fit} with the results from \citet{Nagel2019} for comparison.
We present two models: with planet-planet interactions (PPI) and without planet-planet interactions (noPPI).
In the noPPI model, we find that the eccentricity remains close to zero.
Since, as we show in \autoref{sec:libration}, the librations in eccentricity happen on a similar timescale to the observation time span, we expect that dynamics are important for constraining the eccentricity.
Additionally, we find that in the noPPI model, we converge to nearly the same parameter as \citet{Nagel2019} as in \autoref{tab:best_fit}.
Our results for the PPI model differ most significantly in that we get eccentricities that are of order $0.1$.
Corner plots of the posteriors are presented in Appendix~\ref{sec:posteriors}.

\autoref{fig:rv_fit} shows the nominal best fit solution (red line) with some random samples from the posterior (grey lines) for the phase folded RV data.
The data is phase folded with period $13.347 \unit{days}$ which is the period with the highest power in the RV periodogram as found in \citet{Nagel2019}.
The residuals from the best fit (red) are shown below the RV signal.

\begin{table}
\caption{
    Nominal best fit orbital parameters for GJ4276. 
    PPI/noPPI columns show fitted parameters for the respective model.
    The last column shows the results from \citet{Nagel2019} for reference.
}
\label{tab:best_fit}
\begin{tabular}{lcccc}
\hline
    Orbital Parameter & PPI & noPPI & Nagel (2019) \\ \hline
    $m_1 \sin i \unit{[m_{\earth}]}$ & $15.3 \substack{+0.7 \\ -1.2}$ & $15.72 \substack{+ 3.21\\ -2.97}$ & $15.58 \substack{+0.93 \\ -0.9}$\\
    $m_2 \sin i \unit{[m_{\earth}]}$ & $4.3 \substack{+0.4 \\ -0.6}$ & $4.45 \substack{+0.91 \\ -0.84}$ & $4.40 \substack{+0.44 \\ -0.44}$\\
    $a_1 \unit{[AU]}$ & $0.082 \substack{+0.003 \\ -0.014}$ & $0.082 \substack{+0.002 \\ - 0.002}$ & $0.082 \substack{+0.002 \\ -0.002}$\\
    $a_2 \unit{[AU]}$ & $0.052 \substack{+0.005 \\ -0.002}$ & $0.051 \substack{+0.001 \\ -0.001}$ & $0.051 \substack{+0.001 \\ -0.001}$\\
    $e_1$ & $0.043 \substack{+0.008 \\ -0.002}$ & $\sim 0$ & 0 (fixed) \\
    $e_2$ & $0.070 \substack{+0.004 \\ -0.009}$ & $\sim 0$ & 0 (fixed)\\
    $\omega_1 \unit{[deg]}$ & $45.03 \substack{+5.04 \\ -1.78}$ & $\sim 90$ & 90 (fixed) \\
    $\omega_2 \unit{[deg]}$ & $44.63 \substack{+2.46 \\ -5.56}$ & $\sim 90$ & 90 (fixed) \\
    $\tau_1 \unit{[d]}$ & $0.10 \substack{+0.01 \\ -0.02}$ & $0.10 \substack{+0.02 \\ -0.02}$ & $0.1 \substack{+0.17 \\ -0.17}$\\
    $\tau_2 \unit{[d]}$ & $2.33 \substack{+0.23 \\ -0.33}$ & $2.37 \substack{+0.47 \\ -0.48}$ & $2.35 \substack{+0.18 \\ -0.18}$ \\
    $M_{\star} \unit{[M_{\sun}]}$ & $0.410 \substack{+0.034 \\ -0.020}$ & $0.406 \substack{+0.08 \\ -0.08}$ & $0.406 \substack{+0.030 \\ -0.030}$ \\
    $\gamma \unit{[ms^{-1}]}$ & $0.384 \substack{+0.037 \\ -0.089}$ & $0.39 \substack{+0.08 \\ -0.07}$ & $0.39 \substack{0.18 \\ -0.17}$\\
    $\sigma_{\rm jitter} \unit{[ms^{-1}]}$ & $1.90 \substack{+0.12 \\ -0.15}$ & $1.86 \substack{+0.39 \\ -0.34}$ & $1.89 \substack{+0.18 \\ -0.17}$\\
\hline
\end{tabular}
\end{table}

\subsection{Effect of Dynamics}

To best model the orbital parameters of the system, we need to include the planet-planet interactions.
Even though these interactions may be small compared to the effect of the star, they can add up over long timescales and dramatically change the structure of the planetary system \citep{Batygin2013}.
Including the interactions allows us to better constrain parameters that are dependent on these interactions such as eccentricity.

\begin{figure}
    \centering
    \includegraphics[width=\columnwidth]{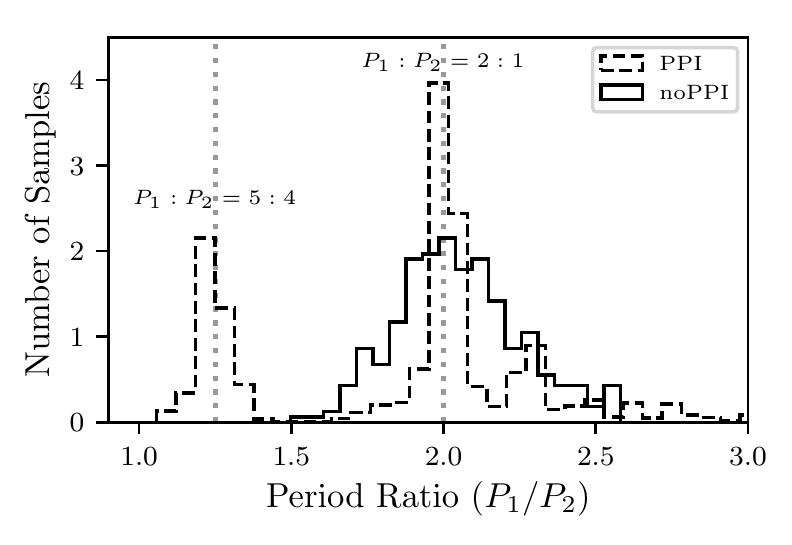}
    \caption{
    Normalized histogram of period ratios in the posterior.
    When dynamical interactions are allowed, we see that the predicted resonance is degenerate, there are peaks in the period ratio at 5:4 and 2:1.
    }
    \label{fig:p_ratio}
\end{figure}

We find that when the planets are allowed to interact, there is a possible solution involving a 5:4 mean motion resonance as well as the originally predicted 2:1 mean motion resonance (\autoref{fig:p_ratio}).
Given the importance of mean motion resonances in planet formation theory \citep{Wang2017}, particularly first order resonances (resonances of the type $(p + 1):p$), the resonant configurations of this system are of dynamical interest.

To explain the 5:4 resonance dynamically, we want to show that planet-planet interactions are relevant even for the short observational window.
In \autoref{sec:libration} we derive an expression for the fast libration timescale, $P_{lf}$, which approximates the interaction timescale between the planets for a timescale shorter than the precession time, and estimate $P_{lf} \approx 2\,000 \unit{days}$.
From numerical results, the change of amplitude of the eccentricity of the inner planet oscillates by almost $0.1$.
Since the observational time span last about half a libration, the significant change in the eccentricity throughout the observations should be apparent in the data.

A fully dynamical fit is important to be able to predict the eccentricity and, eventually (with enough follow up observations) the inclination.
In our dynamical model, the eccentricities are of order $\sim 0.1$, large enough that they should not be assumed to be 0.
This is in line with typical eccentricities of 2 planet systems around M-stars (ie \citet{Limbach2015}).

\section{Resonance Analysis} \label{sec:eqns}

\subsection{Basic Equations}

Here we define the equations and notation used to analyze the two planet resonance analytically.
We follow closely the work of \cite{Rein2009}.
We start from Hamiltonian formalism using Jacobi Heliocentric coordinates \citep{Sinclair1975}.
Here the positions  of the planets, with reduced mass $m_i$ are given by $\mathbf{r_i}$ as measured from the central mass.
In our case, $i = 1, 2$ where $1, 2$ refer to the outer and inner planet respectively.

The general Hamiltonian for this system, correct to second order in the masses is given by
\begin{equation}
\begin{split}
    H &= \frac{1}{2} (m_1 |\dot{\mathbf{r_1}}|^2 + m_2 |\dot{\mathbf{r_2}}|^2)
    - \frac{\mathcal{G}M_1m_1}{|\mathbf{r_1}|} - \frac{\mathcal{G}M_2m_2}{|\mathbf{r_2}|} \\ 
    & - \frac{\mathcal{G}m_1 m_2}{|\mathbf{r_{12}}|} 
    + \frac{\mathcal{G}m_1m_2 \mathbf{r_1} \cdot \mathrm{\mathbf{r_2}}}{|\mathbf{r_1}|^3} 
    \label{eq:hamiltonian}
\end{split}
\end{equation}
where $M$ is the stellar mass, $M_1 = M + m_1$, $M_2 = M + m_2$ and $\mathbf{r_{12}} = \mathbf{r_2} - \mathbf{r_1}$ and $\mathcal{G}$ is the gravitational constant.
We then seek to express the Hamiltonian in terms of the energies, $E_i= -\mathcal{G}m_iM_i / (2a_i)$ and the resonant angles, $\phi_i$.
For a general $(p + 1): p$ first order resonance, the resonant angles are expressed as
\begin{equation}
    \label{eq:res_angle}
    \begin{split}
    \phi_1 &= (p+1) \lambda_1 - p\lambda_2 - \varpi_2 \\
    \phi_2 &= (p+1)\lambda_1 - p\lambda_2 - \varpi_1
    \end{split}
\end{equation}
where $\lambda_i$ is the mean longitude of the $i$-th planet and $\varpi_i$ the longitude of pericenter. 
In our case, $p = 1$ since we are considering the 2:1 mean motion resonance.

We follow the method of \citet{Brouwer1961}, which expresses the Hamiltonian as a Fourier expansion in terms of the resonant angles $\phi_1, \phi_2$.
We expect that near a first order resonance, the angles $\phi_1$, $\phi_2$ will vary slowly.
Following the procedure of \citet{Papaloizou2005}, we can keep only terms which are a linear combination of $\phi_1$ and $\phi_2$. 
We then separate the Hamiltonian into the Keplerian part and the disturbing part as $H = E_1 + E_1 + H_{12}$.
The disturbing part, $H_{12}$ is given in terms of the resonant angles $\phi_i$, the eccentricities $e_i$, the masses $m_i$ and $\alpha = a_1 / a_2$ as

\begin{equation}
    \label{eq:disturbing}
    H_{12} = \frac{-\mathcal{G} m_1 m_2}{a_1}
    \sum C_{k,l}\left(\alpha, e_1, e_2\right) \cos(l \phi_1 + k \phi_2)
\end{equation}
where the constant $C_{k, l}$, for $k, l$ any integers, depends only on $\alpha, e_i$.
Expansions for $C_{k,l}$ can be found in the appendix of many references on Celestial Mechanics such as \citet{Murray2000}.

\begin{figure}
    \centering
    \includegraphics[width=\linewidth]{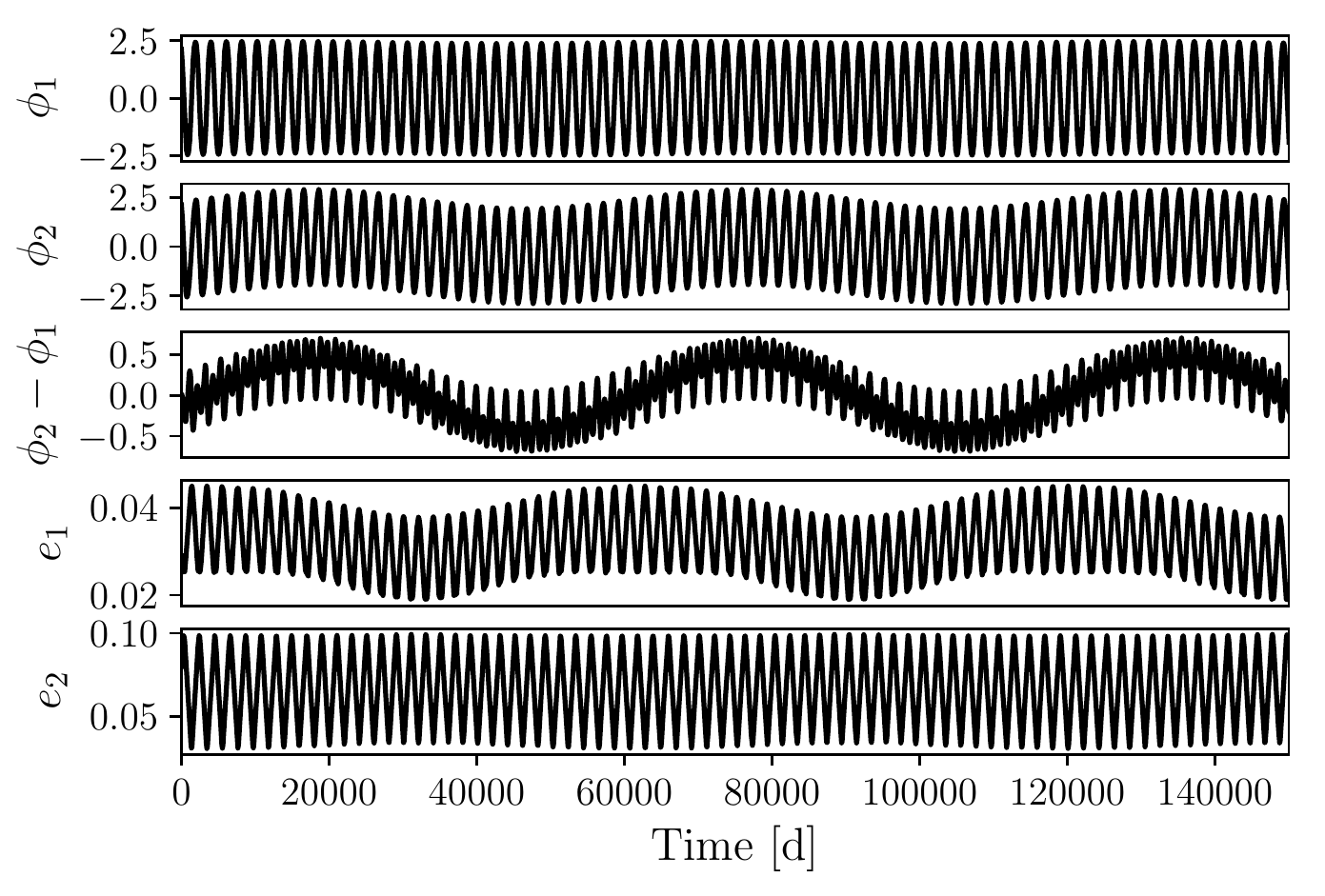}
    \caption{
    Time evolution of the resonant angles (in radians) and the eccentricities for the system GJ4276.
    The dominance of the fast mode in the oscillations of $\phi_1, \phi_2$ with period $\sim 2\,000 \unit{days}$ and the slow mode in the oscillations of $\phi_2 - \phi_1$ with period $\sim 60\,000 \unit{days}$ can be seen.
    }
    \label{fig:eccentricity}
\end{figure}

Then we solve the equations of motion for the angles, $\phi_1, \phi_2$ and their difference $\zeta = \phi_2 - \phi_1$, as in \citet{Rein2009}.
\autoref{fig:eccentricity} shows the librations in the eccentricities and resonant angles from a numerical integration over $1.5 \times 10^5 \unit{days}$.
The top two panels show the libration in radians of the resonant angles (\autoref{eq:res_angle}) where we see the librations are dominated by a fast mode.
We also see that the fast librations in $\phi_1, \phi_2$ are approximately in phase with the same period.
The middle panel shows the librations of $\zeta$ in radians.
The librations of $\zeta$ are dominated by the slow mode (which we see corresponds to the slow libration in $\phi_2$).
The slow mode is much longer than the fast mode so the modes can be decoupled.
The bottom two panels show the librations in the eccentricities themselves.
We see that they are dominated by the fast mode but we can see the slow mode in the librations of $e_1$. 
The change in amplitude of the eccentricities of one oscillation is fairly large, which would be the primary effect we would see during observation.

\subsection{Fast Libration Mode} \label{sec:libration}

Following \citet{Rein2009}, the linearized equations, to first order in the planetary masses give
\begin{equation}
    \frac{d \phi_i}{dt} \approx (p + 1)n_1 - n_2
\end{equation}
where $n_i$ are the mean motion of the planets.
So in this approximation, $\phi_1 \approx \phi_2$.
Then, to first order in the planet masses, the angles $\phi_i$ evolve according to

\begin{equation}
    \label{eq:fast_de}
    \frac{d^2 \phi_i}{dt^2} + \omega_{lf}^{2}\phi_i = 0,\quad (i = 1, 2)
\end{equation}
\citet{Rein2009} derive the libration frequency $\omega_{lf}$ as:
\begin{equation}
    \label{eq:wlf}
    \omega_{lf}^2 = -\frac{3 n_{2}^{2} m_{2}}{M}\left(1 + \frac{m_1}{\alpha m_2}\right) \sum C_{k,l}(k+l)^2
\end{equation}
Applying this condition and expanding $C_{k,l}$ and keeping only terms of order unity in eccentricity, we approximate the fast libration time.
We calculate the fast libration period, $P_{lf} \approx 2\,000 \unit{days}$ which agrees with the numerical result.

\subsection{Slow Libration Mode}

For the slow libration mode, we want to look for librations on the timescale that $\zeta = \phi_2 - \phi_1$ changes. 
In this case, it is no longer true that $\phi_1 \approx \phi_2$.
We now seek solutions to oscillations of the form
\begin{equation}
    \label{eq:slow_de}
    \frac{d^2 \zeta}{dt^2} + \omega_{ls}^2\zeta = 0,
\end{equation}
where for a $(p + 1):p$ resonance, using equation 33 from \citet{Rein2009}, we estimate the slow libration timescale $P_{ls}$ is approximately $55\,000 \unit{days}$.

This is a much longer mode of libration which we do not expect to see in the observations.
We will therefore limit discussions to the fast mode.

\subsection{Observational Implications}

Even though the observation window of $\sim 700 \unit{days}$ is fairly short, since the fast librations timescale is $\sim 2\,000 \unit{days}$, we should be able to detect this in the RV data.
With the observational timescale being approximately half the libration time, which expect the eccentricities of the planets to change significantly.
We note that although the librations will affect the fitting process, we would need to measure the RV signal for several libration periods to be able to claim a definite detection of planet-planet interactions.
This could be tested with longer follow up observation over at least several libration periods.
Though this time is fairly long, compared to libration timescales in systems similar to our own solar system it is much shorter, thus making it a potentially viable target in the future.

The oscillations on the slow timescale are not large enough that we would expect to see them without significantly longer observations as the timescale for this mode is over $170\unit{years}$.
It is still useful to know that we might see small changes to the eccentricity, of the outer planet in particular, due to the slow mode.

\section{Stability} \label{sec:stability}

\subsection{Stability Analysis with Inclination}

To study the stability of the system, we focus on regions of parameter space that are occupied by the samples of our posterior.
We integrate a total of $8\,500$ realizations of GJ4276 for an integration time of $10^6 \unit{orbits}$.
We also run a smaller subset of solutions for $10^7 \unit{orbits}$ and find no significant differences in the results.
As with previous numerical simulations, we use the integrator \texttt{IAS15} with the \texttt{REBOUND} package.
We find that after $10^6 \unit{orbits}$, $\sim 97\%$ of realizations are stable.

The $3\%$ of unstable realizations do not show any clear preference for happening near either one of the resonances.
Therefore, we cannot rule out either resonant configuration from stability alone.

To augment the results from the direct numerical integrations, we also measure the mean exponential growth of nearby orbits (MEGNO) \citep{Cincotta2000}.
The MEGNO indicator, $\langle Y \rangle$, provides a way to quickly distinguish chaotic and quasi periodic regions in the phase space of a dynamical system.
To compute $\langle Y \rangle$ we integrate the variational equations \citep{Mikkola1999} using \texttt{IAS15} and the utilities included in \texttt{REBOUND}.
If $\langle Y \rangle = 2$ then this indicates that the system exhibits stable, quasi periodic behaviour.
A value larger than $2$ indicates chaotic behaviour and therefore that the system is likely to become unstable over a shorter timescale.

\begin{figure}
    \centering
    \includegraphics[width=\linewidth]{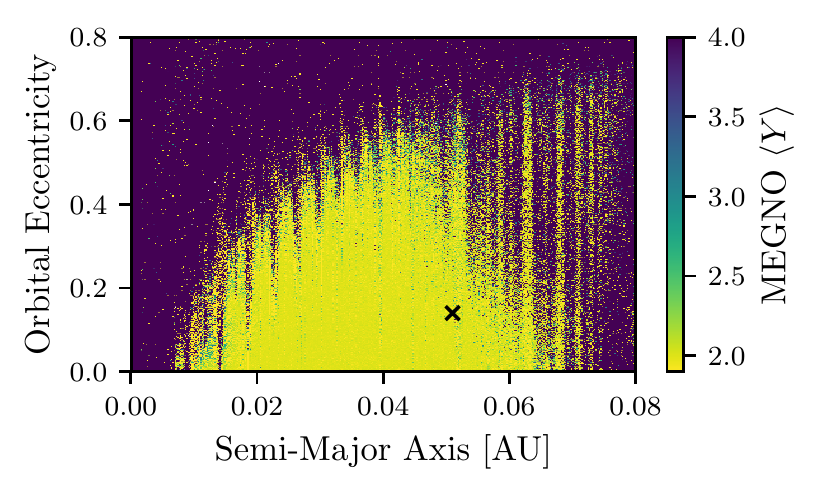}
    \caption{
    MEGNO stability indicator in the region of the best fit solution.
    Bright yellow regions ($\sim 2$) are stable quasi-periodic solutions according to the MEGNO indicator.
    The black cross shows the nominal mean orbital parameters for the inner planet (as listed in \autoref{tab:best_fit}).
    }
    \label{fig:megno}
\end{figure}

A cross section of the parameter space in the plane of the semi-major axis and eccentricity of the inner planet is presented in \autoref{fig:megno} with the color indicating $\langle Y \rangle$.
All of the other orbital parameters remain as in \autoref{tab:best_fit}.
The black cross indicates the location of the nominal mean orbital parameters, which is well within the island of stability. 

\subsection{Effects of Including Inclination}

\begin{figure}
    \centering
    \includegraphics[width=\linewidth]{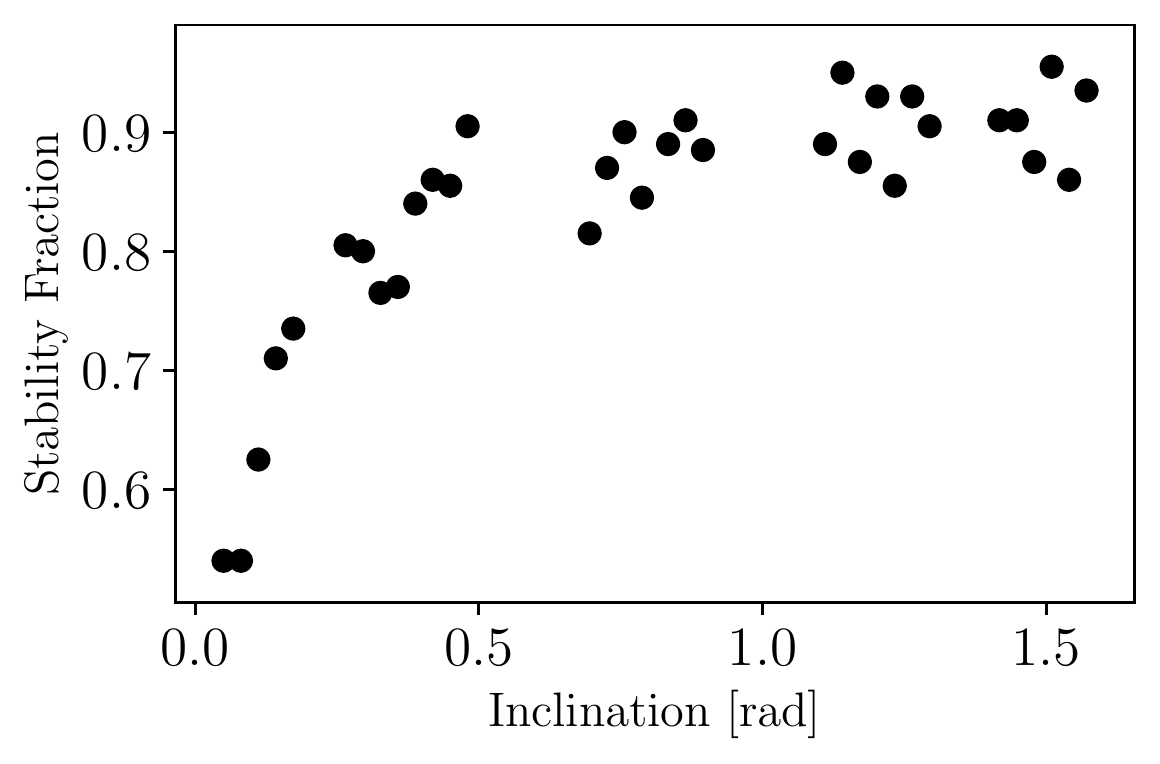}
    \caption{
    The stability fraction of the system for different values of inclination. 
    For each value of $i$, 500 samples were drawn randomly from the posterior and integrated for $10^6$ orbits.
    The stability fraction is the fraction of realizations which did not have close encounters.
    }
    \label{fig:inc-stability}
\end{figure}

The inclination of the system effects the masses so we can use stability to place a bound on the true masses, $m_{\rm true} = m_{\rm min} / \sin i$.
To do this, we draw 500 samples out of the posterior distribution for 100 different values of inclination and simulate the system for $10^6$ orbits of the inner planet.
We can then calculate the fraction of simulations which do not have close encounters during this time.

\autoref{fig:inc-stability} shows the stability fraction as a function of inclination.
The stability fraction starts to rapidly decrease around $i \lessapprox 0.5\unit{rad}$, which forms a rough lower constraint on the inclination of the system.
Using this value, we can also place a constraint on the maximal masses of the two planets to be $m_{1} \lessapprox 32 \unit{m_{\earth}}$ and $m_{2} \lessapprox 9 \unit{m_{\earth}}$.
We also test the effect of giving the planets a small mutual inclination of $\sim 3^\circ$ and find this has no significant effect on the results.

The constraints on the inclination are very loose.
Having tighter constraints on the orbital parameters would allow us to better estimate the stability fraction and therefore the inclination as well.

\section{Discussion and Conclusion} \label{sec:discussion}

In this study, we reanalyzed the 100 RV measurements for the system GJ4276.
We focused on the possibility of fitting a two planet model to the data.
In particular, in contrast to previous studies, we performed a fully dynamical fit where the planets were allowed to interact with one another.
To fit the model, we simulated the RV curve using numerical simulations with \texttt{REBOUND} while sampling the posterior with the Markov Chain Monte Carlo sampler \texttt{EMCEE}.

The best fit two planet model we predict has an outer planet with minimal mass $\sim 15.3 \unit{m_{\earth}}$ and semi-major axis $\sim 0.082 \unit{AU}$ and an inner planet with minimal mass $\sim 4.3 \unit{m_{\earth}}$ and semi-major axis $0.052 \unit{AU}$.
We find that the eccentricities of the inner and outer planet respectively are $\sim 0.07$ and $\sim 0.043$.
This is consistent with typical eccentricity values for planets around M-stars \citep{Limbach2015}.
\citet{Nagel2019} estimate the tidal circularization time for the Neptune mass planet to be at least $10\unit{Gyr}$, supporting the idea that the eccentricities are not necessarily 0.
Additionally, we place a rough constraint of $i \lessapprox 0.5\unit{rad}$ on the inclination.
This provides an estimate for the maximum masses of the planets of $m_1 \lessapprox 32 \unit{m_{\earth}}$ and $m_2 \lessapprox 9 \unit{m_{\earth}}$.

One argument in favour of the two planet model over the single eccentric model proposed in \citet{Nagel2019} is that the eccentricity required for the one planet model is $\approx 0.37$.
There have only been 13 confirmed detections of exoplanets around M-stars with $e > 0.3$ out of $73$ confirmed observations with eccentricity values.
The eccentricity predicted for the one planet model is comparable to the recently published planet Gl96b with $e \approx 0.44$ \citep{Hobson2018}.

From \autoref{fig:p_ratio} we can see that in the PPI model in particular, the system favours configurations which are in a first order mean motion resonance.
The existence of any resonance suggests that the planets got into this configuration through migration in the protoplanetary disk \citep{Correia2017}.
Which resonance the system favours can give us insight into how the migration took place.
Since the 5:4 resonance places the planets closer together than the 2:1 resonance, the planets would likely pass through the 2:1 resonance during migration.
This would predict a faster migration rate and therefore larger disk.
We slightly favour a 2:1 mean motion resonance over a configuration in a 5:4 mean motion resonance.

We can also infer that since migration is likely, the planets formed prior to disk dispersal.
This means that they would likely have had time to accrete an atmosphere.
However, both planets are well inward of the habitable zone which has inner edge at $\sim 0.115\unit{AU}$ \citep{Kopparapu2013}.

The system has a variety of interesting dynamical features that merit further study of its origin and evolution.
If a second planet is confirmed, it would be only the third discovered M-dwarf host with two planets in a 2:1 mean motion resonance.
It would also be the first of these systems to have a 2:1 resonance without a resonant chain (GJ876 is a 4:2:1 and TRAPPIST-1 a 8:5:3:2:1 resonant chain).
If it turns out that there is not a second planet and it is a highly eccentric Neptune mass planet, then it would be one of the most eccentric planets around an M-dwarf with such a short orbital period.
Given that GJ4276 shows such interesting orbital configurations, it is well worth considering for follow up observations.
Given the libration time we predict, observing for 5 years would allow us to determine whether or not there is a second planet and put constraints on the inclination of the system.

\section*{Acknowledgements}

We thank the CARMENES group for making the RV data for GJ4276 publicly available.
This research has been supported by the NSERC Discovery Grant RGPIN-2014-04553.
This research made use of the software packages \texttt{REBOUND} \citep{ReinLiu2012}, \texttt{EMCEE} \citep{Foreman2013}, \texttt{matplotlib} \citep{Hunter2007} and \texttt{corner.py} \citep{corner}.


\bibliographystyle{mnras}
\bibliography{bibliography} % bibtex with file bibliography.bib

%%%%%%%%%%%%%%%%% APPENDICES %%%%%%%%%%%%%%%%%%%%%

\appendix
\section{Corner Plots for MCMC Posteriors}\label{sec:posteriors}

\begin{figure*}
    \centering
    \includegraphics[width=\linewidth]{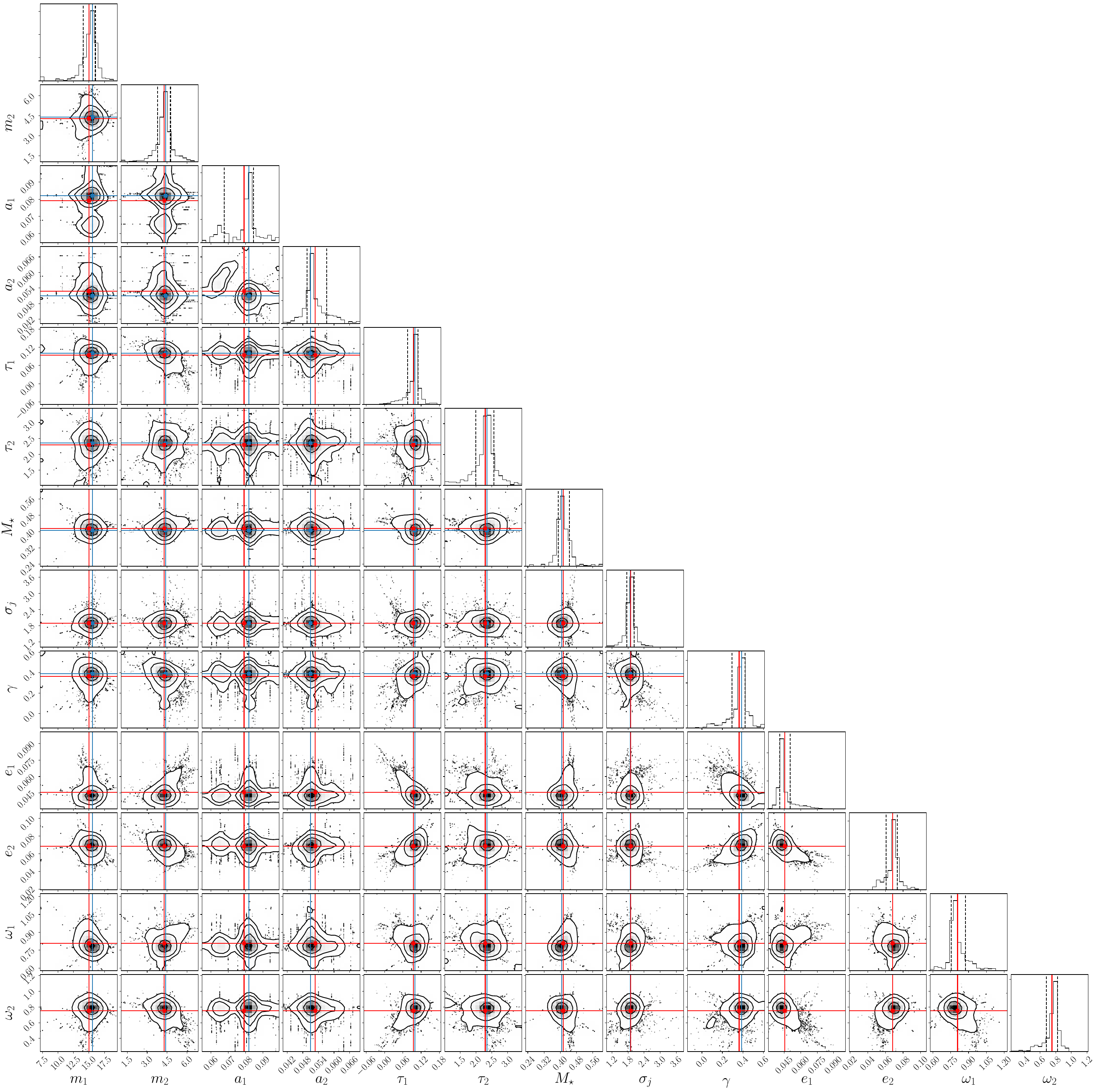}
    \caption{Two dimensional projection of the posterior probability distribution sampled using MCMC for the PPI model. Dashed lines on the histograms indicate the $1\sigma$ confidence interval.}
    \label{fig:corner_int}
\end{figure*}

\begin{figure*}
    \centering
    \includegraphics[width=\linewidth]{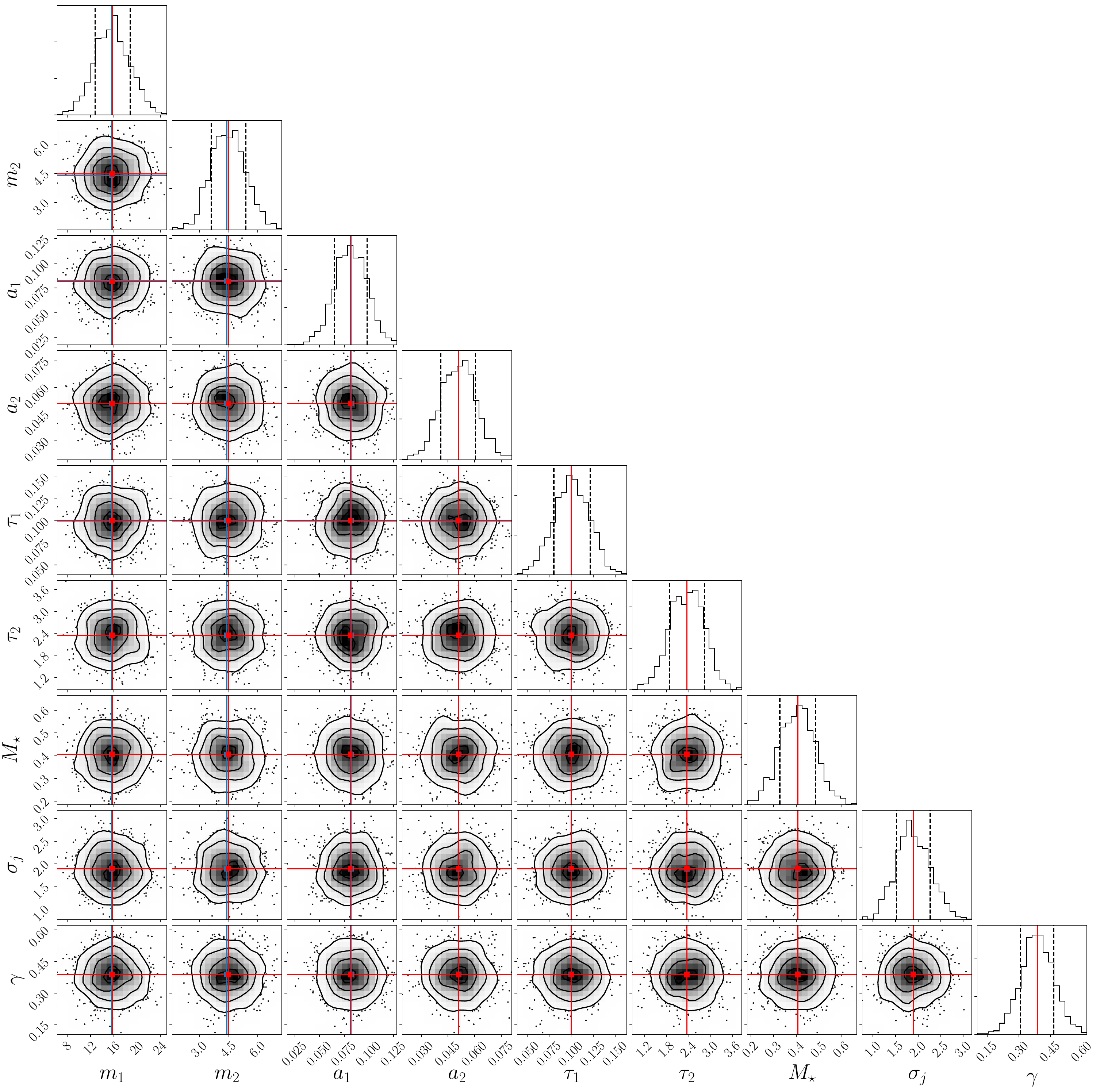}
    \caption{Same as \autoref{fig:corner_int} but for the noPPI model.}
    \label{fig:corner_non_int}
\end{figure*}

Figures~\ref{fig:corner_int} and~\ref{fig:corner_non_int} show corner plots for the posteriors from the MCMC sampler for the PPI and noPPI models, respectively.
The masses $m_1, m_2$ are the minimal masses ($m_1 \sin i$, $m_2 \sin i$).
The blue lines show the nominal predicted values from \citet{Nagel2019} and the red lines the nominal best fit values from this work.
The dashed lines in the histograms show the $1\sigma$ confidence interval.
Above each histogram, the mean value for the parameter is listed with uncertainty given by the bounds of the $1\sigma$ confidence interval. 
The values are repeated in \autoref{tab:best_fit} for convenience.

We note that when interactions are allowed, the parameters are more restricted and the distributions have sharper peaks as well as and non-Gaussian distributions.
In  the noPPI model, the distributions are essentially Gaussian and very similar to the results from \citet{Nagel2019}.
Since in the noPPI case the eccentricities are negligibly small, we exclude the parameters from the corner plot.

%%%%%%%%%%%%%%%%%%%%%%%%%%%%%%%%%%%%%%%%%%%%%%%%%%

% Don't change these lines
\bsp	% typesetting comment
\label{lastpage}
\end{document}